\documentclass[twocolumn,aps,showpacs,preprintnumbers,amsmath,amssymb, superscriptaddress]{revtex4-1}
\pdfoutput=1

\usepackage{bm}
\usepackage{graphicx}
\usepackage{color}
\usepackage{xspace}

\renewcommand{\vec}{\mathbf}

\begin{document}

\title{Experimental investigation of the electronic structure of Ca$_{0.83}$La$_{0.17}$Fe$_2$As$_2$}

\author{Y.-B. Huang}
\affiliation{Beijing National Laboratory for Condensed Matter Physics, and Institute of Physics, Chinese Academy of Sciences, Beijing 100190, China}
\author{P. Richard}\email{p.richard@iphy.ac.cn}
\affiliation{Beijing National Laboratory for Condensed Matter Physics, and Institute of Physics, Chinese Academy of Sciences, Beijing 100190, China}
\author{J.-H. Wang}
\affiliation{Texas Center for Superconductivity and Department of Physics, University of Houston, TX 77204-5002, USA}
\author{X.-P. Wang}
\affiliation{Beijing National Laboratory for Condensed Matter Physics, and Institute of Physics, Chinese Academy of Sciences, Beijing 100190, China}
\affiliation{Paul Scherrer Institut, Swiss Light Source, CH-5232 Villigen PSI, Switzerland}
\author{X. Shi}
\affiliation{Beijing National Laboratory for Condensed Matter Physics, and Institute of Physics, Chinese Academy of Sciences, Beijing 100190, China}
\affiliation{Paul Scherrer Institut, Swiss Light Source, CH-5232 Villigen PSI, Switzerland}
\author{N. Xu}
\affiliation{Beijing National Laboratory for Condensed Matter Physics, and Institute of Physics, Chinese Academy of Sciences, Beijing 100190, China}
\author{Z. Wu}
\affiliation{Texas Center for Superconductivity and Department of Physics, University of Houston, TX 77204-5002, USA}
\author{A. Li}
\affiliation{Texas Center for Superconductivity and Department of Physics, University of Houston, TX 77204-5002, USA}
\author{J.-X. Yin}
\affiliation{Beijing National Laboratory for Condensed Matter Physics, and Institute of Physics, Chinese Academy of Sciences, Beijing 100190, China}
\author{T. Qian}
\affiliation{Beijing National Laboratory for Condensed Matter Physics, and Institute of Physics, Chinese Academy of Sciences, Beijing 100190, China}
\author{B. Lv}
\affiliation{Texas Center for Superconductivity and Department of Physics, University of Houston, TX 77204-5002, USA}
\author{C. W. Chu}
\affiliation{Texas Center for Superconductivity and Department of Physics, University of Houston, TX 77204-5002, USA}
\author{S. H. Pan}
\affiliation{Texas Center for Superconductivity and Department of Physics, University of Houston, TX 77204-5002, USA}
\affiliation{Beijing National Laboratory for Condensed Matter Physics, and Institute of Physics, Chinese Academy of Sciences, Beijing 100190, China}
\author{M. Shi}
\affiliation{Paul Scherrer Institut, Swiss Light Source, CH-5232 Villigen PSI, Switzerland}
\author{H. Ding}\email{dingh@iphy.ac.cn}
\affiliation{Beijing National Laboratory for Condensed Matter Physics, and Institute of Physics, Chinese Academy of Sciences, Beijing 100190, China}

\date{\today}

\begin{abstract}
We performed a combined angle-resolved photoemission spectroscopy and scanning tunneling microscopy study of the electronic structure of electron-doped Ca$_{0.83}$La$_{0.17}$Fe$_2$As$_2$. A surface reconstruction associated with the dimerization of  As atoms is observed directly in the real space, as well as the consequent band folding in the momentum space. Besides this band folding effect, the Fermi surface topology of this material is similar to that reported previously for BaFe$_{1.85}$Co$_{0.15}$As$_2$, with $\Gamma$-centred hole pockets quasi-nested to M-centred electron pockets by the antiferromagnetic wave vector. Although no superconducting gap is observed by ARPES possibly due to low superconducting volume fraction, a gap-like density of states depression of $7.7\pm 2.9$ meV is determined by scanning tunneling microscopy.
\end{abstract}



\maketitle

The parent compound of the AFe$_2$As$_2$ (A = Ca, Sr, Ba) family of Fe-based superconductors is an antiferromagnetic (AFM) bad metal characterized by a Dirac cone electronic dispersion \cite{RichardPRL2010}. Superconductivity occurs in this material after suppression of long-range AFM order following the application of pressure \cite{Kimber,Torikachvili2} or the injection of electronic carriers into the Fe-As layers \cite{GF_Chen2,Rotter_PRL2008}. Recently, a few groups reported the synthesis of rare-earth-doped CaFe$_2$As$_2$ samples exhibiting superconductivity \cite{LvPNAS2011,Gao_EPL95,Saha_PRB85} at temperature as high as 49 K, raising enthusiasm in the community as well as fundamental questions. In addition to carrier doping, the various size of the substituted rare-earth can act as chemical pressure, thus allowing a high flexibility in shaping the electronic structure of these materials. However, these new compounds show only low superconducting (SC) volume fractions, which may suggest a minor SC phase rather different from the bulk SC phase detected in other doped AFe$_2$As$_2$ compounds with a critical temperature $T_c$ below 38 K. Moreover, such a high $T_c$ is much higher than what one would expect from a quasi-nesting \cite{RichardRoPP2011} approach based on the calculation of the Lindhard function at the M($\pi,\pi$) point in the electron-doped side of the phase diagram as compared to the hole-doped side \cite{Neupane_PRB2011}. To address these issues, it is imperative to characterize the electronic structure of these new superconductors in the momentum and real spaces.

Angle-resolved photoemission spectroscopy (ARPES) \cite{RichardRoPP2011} and scanning tunneling microscopy (STM) \cite{HoffmanRoPP2011} are particularly suitable to probe the electronic structure and the SC gap of both electron-doped and hole-doped AFe$_2$As$_2$ compounds in the momentum and real spaces, respectively. In this Letter, we report a combined ARPES and STM investigation of the electronic structure of Ca$_{0.83}$La$_{0.17}$Fe$_2$As$_2$. Although clear electronic band dispersion and Fermi surface (FS) are revealed by ARPES, the SC gap cannot be resolved in the momentum space. Despite poorly defined spectral features likely due to a small SC volume fraction, a characteristic gap-like density of states depression of $\sim 8$ meV is derived from the STM data. In addition, a $a\times 2a$ surface reconstruction corresponding to a dimerization of As atoms is observed from topographic images, which leads to a band folding detected in the ARPES data. 

Single-crystals with composition Ca$_{0.83}$La$_{0.17}$Fe$_2$As$_2$ determined from wavelength-dispersive spectrometry chemical analysis were grown by the self-flux method \cite{LvPNAS2011} and their $T_c$ was determined to be up to 42 K with a second transition at 20 K from resistivity measurements. ARPES measurements were performed at Swiss Light Source beamline SIS using a VG-Scienta R4000 electron analyzer with photon energy ranging from 26 to 66 eV. The angular resolution was set to 0.2$^{\circ}$ and the energy resolution to 8-20 meV. All samples were cleaved \emph{in situ} and measured at 11.5 K in a working vacuum better than 5$\times$10$^{-11}$ torr. The STM measurements were performed at the Texas Center for Superconductivity at the University of Houston using a homemade STM equipped with a polycrystalline tungsten tip. Samples were cleaved \emph{in situ} below 30 K and cooled down to 4.2 K for measurements under cryogenic vacuum. For consistency between the STM and ARPES descriptions of the experimental results, we label the momentum values with respect to the 2 Fe/unit cell Brillouin zone (BZ).

\begin{figure}[!t]
\begin{center}
\includegraphics[width=8.5cm]{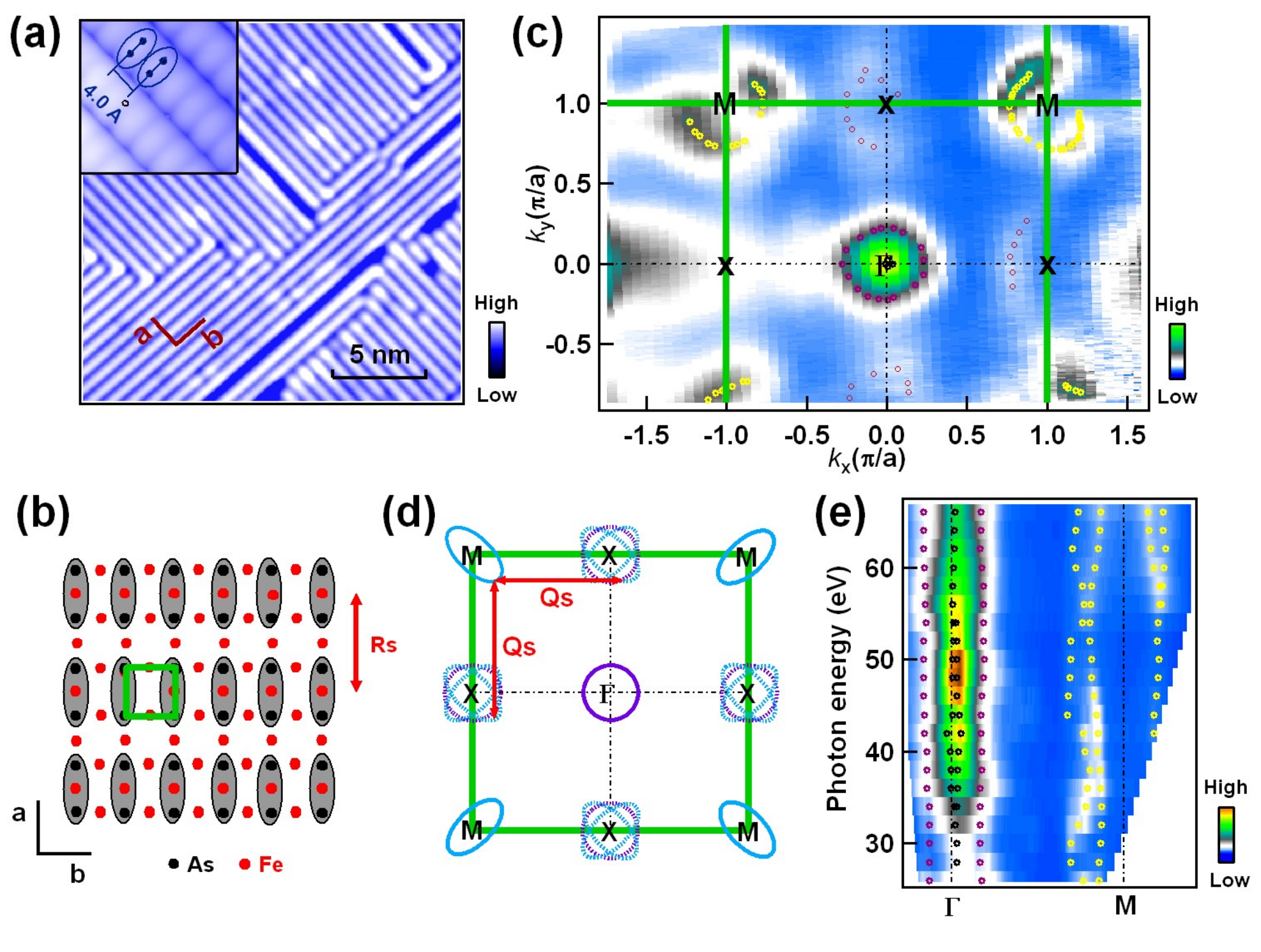}
\end{center}
\caption{\label{Figure1_FS}(Colour online) (a) STM topographic map ($20 \times 20$ nm$^2$) of a Ca$_{0.83}$La$_{0.17}$Fe$_2$As$_2$ sample recorded with a bias of -100 mV. A zoom is given in inset where dimerization of As atoms is observed. (b) Schematic representation of the As dimerization. The green square corresponds to the 2 Fe/unit cell. (c) FS intensity plot ($\pm 10$ meV integration at $E_F$) of Ca$_{0.83}$La$_{0.17}$Fe$_2$As$_2$ recorded at 63 eV with circular-right polarization. Extracted FS crossings are indicated by symbols. (d) Schematic representation of the $\vec{Q_s}$ electronic band folding using the 2 Fe/unit cell notation. The BZ delimited in green is consistent with the unit cell defined in (b). (e) Photon energy dependence of the near-$E_F$ photoemission intensity ($\pm 10$ meV) along $\Gamma$-M. Extracted FS crossings are indicated by symbols.}
\end{figure}

We show in Fig. \ref{Figure1_FS}(a) a topographic image ($20\times 20$ nm$^2$) of the surface of a Ca$_{0.83}$La$_{0.17}$Fe$_2$As$_2$ sample. At this scale, we observe fine lines along either the orthogonal a- or b- axes, indicating a local breakdown of the 4-fold symmetry at the surface. Zooming further (inset of Fig. \ref{Figure1_FS}(a)), we distinguish 4 \AA\xspace long features that are ascribed to As dimers also spaced by 4 \AA\xspace along dimer rows, as emphasized by the ellipses drawn in inset and sketch in Fig. \ref{Figure1_FS}(b). Each dimer row is spaced from its neighbours by a distance $R_s$ twice as large. To check whether such strong surface reconstruction ($a\times 2a$) affects the Fe electronic states, we recorded a FS mapping using ARPES. The result, displayed in Fig. \ref{Figure1_FS}(c), shows relatively strong intensity at the $\Gamma (0,0)$ and M$(\pi,\pi)$ points, as observed in other Fe-pnictides \cite{RichardRoPP2011}. In addition, non-negligible intensity is detected at the X$(\pi,0)$ point. As illustrated schematically in Fig. \ref{Figure1_FS}(d), this additional intensity is consistent with the folding by the wave vector $\vec{Q_s}$ of bands that would be induced from the dimerization of As atoms at the surface. We mention that the ARPES FS pattern is 4-fold symmetric rather than 2-fold symmetric due to the average of the twin domains observed by STM (Fig. \ref{Figure1_FS}(a)). 

\begin{figure}[!t]
\begin{center}
\includegraphics[width=8.5cm]{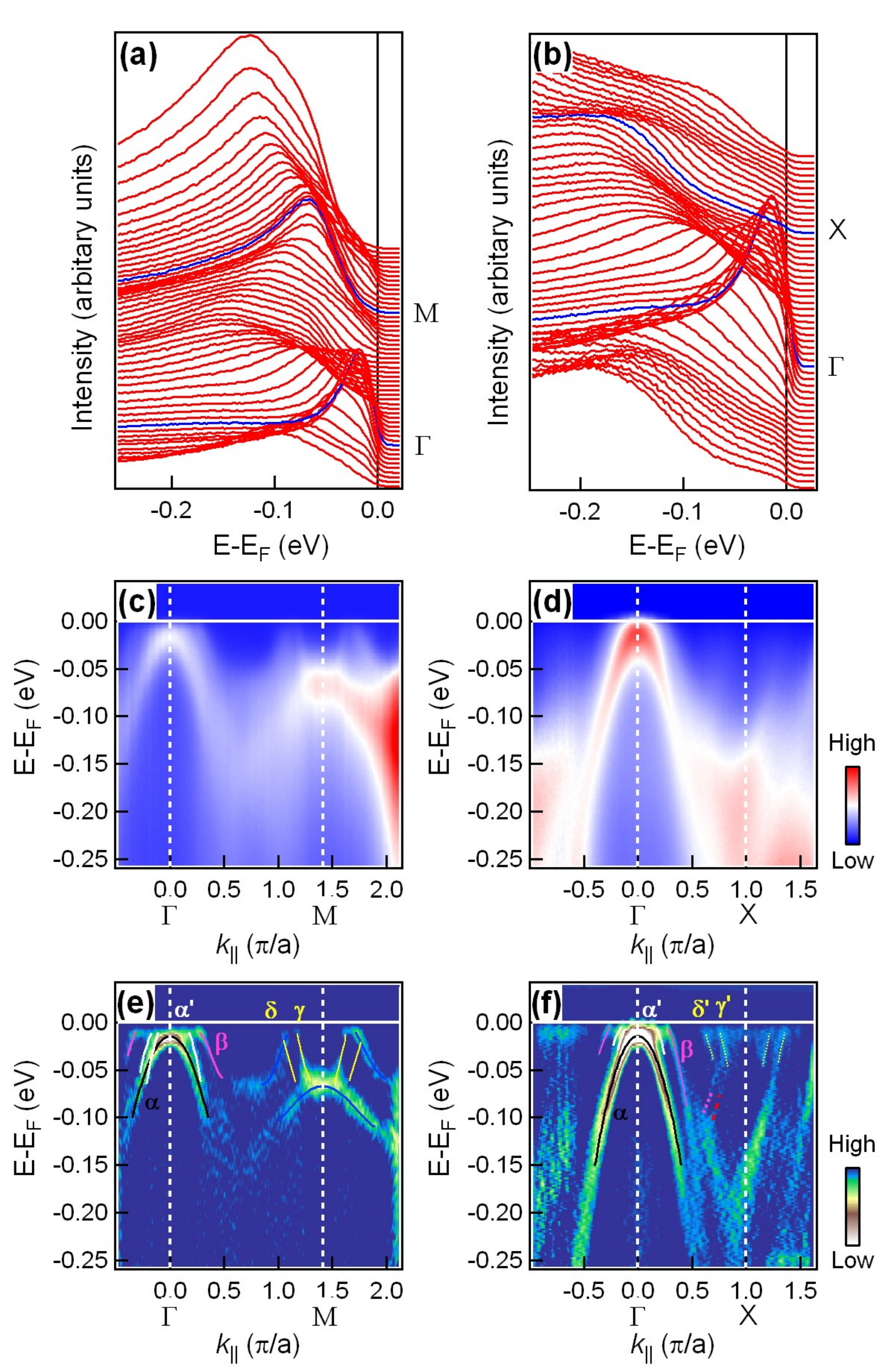}
\end{center}
\caption{\label{Figure2_structure}(Colour online) In-plane electronic band dispersion recorded by ARPES with 63 eV photons along high-symmetry lines. (a)-(b) EDCs. (c)-(d) Corresponding ARPES intensity plots. (e)-(f) Corresponding 2D curvature intensity plots \cite{P_Zhang_RSI2011}. Solid lines are guides to the eye for the main band dispersions whereas dashed lines refer to folded bands.}
\end{figure}

We display in Fig. \ref{Figure1_FS}(e) the photon energy dependence of the ARPES intensity near the Fermi level ($E_F$) along $\Gamma$-M. Recalling that the photon energy dependence is directly related to the momentum dispersion perpendicular to the probed surface ($k_z$) \cite{DamascelliPScrypta2004}, the data indicate weak $k_z$ dispersion. The in-plane band dispersion measured at 63 eV along $\Gamma$-M and $\Gamma$-X, is given in Fig. \ref{Figure2_structure}. The energy distribution curves (EDCs) shown in the top row, the corresponding photoemission intensity plots displayed in the middle row and the 2D curvature intensity plots given in the bottom row indicate clear band dispersion, as expected for a single crystalline phase. We distinguish 3 hole bands around the $\Gamma$ point, labeled $\alpha$, $\alpha^{\prime}$ and $\beta$ in Figs. \ref{Figure2_structure}(e) and \ref{Figure2_structure}(f). As with BaFe$_{1.85}$Co$_{0.15}$As$_2$ \cite{Terashima_PNAS2009}, the former sinks slightly below $E_F$ whereas the other bands contribute to the FS. Yet, the top of the $\alpha$ band is mainly responsible for the intensity integrated near $E_F$ in Fig. \ref{Figure1_FS}(c). At the M point, we can resolve 2 electron bands named $\gamma$ and $\delta$. Their size is qualitatively similar to that of the $\beta$ FS, indicating the possibility of electron-hole quasi-nesting. We also point out that the system possibly shows imprints of the antiferromagnetic ordering characteristic of the parent compound. In fact, $\Gamma$ to M band folding has been evidenced in CaFe$_2$As$_2$ \cite{LiuPRL2009}, SrFe$_2$As$_2$ \cite{Y_Zhang_PRL102} and BaFe$_2$As$_2$ \cite{RichardPRL2010}. Although with much less spectral intensity than in these parent compounds, Figs. \ref{Figure2_structure}(c)-(f) show the folding of the $\beta$ band in the M point region.  

In agreement with the FS mapping, weak but finite intensity is detected around the X point. Similarly to the main bands, the 2D curvature intensity plot shown in Fig. \ref{Figure2_structure}(f) allows a better visualization of the corresponding dispersion bands. Hence, two electron bands, namely $\gamma^{\prime}$ and $\delta^{\prime}$, are centred at X. Their Fermi velocities and FS sizes are in good correspondence with the M-centred $\gamma$ and $\delta$ bands, suggesting that they result from a M to X band folding. As discussed above, this observation is supported by our STM observation of surface reconstruction induced by the dimerization of the As atoms. This scenario implies also a $\Gamma$ to X band folding. Albeit with much weaker intensity, the 2D curvature intensity plot shown in Fig. \ref{Figure2_structure}(f) also exhibits hole band dispersions, as emphasized by purple dashed lines. 

\begin{figure}[!t]
\begin{center}
\includegraphics[width=8.5cm]{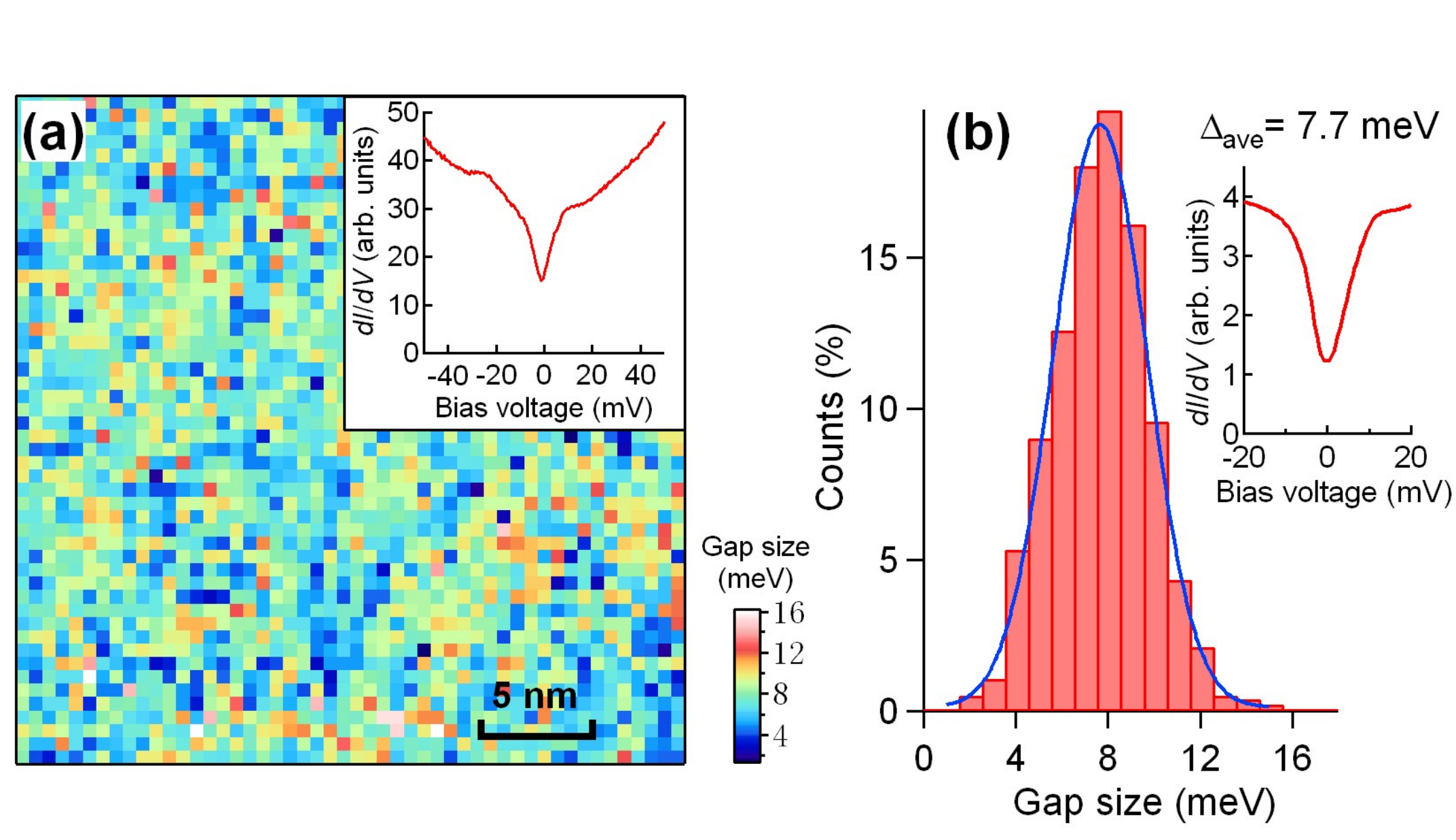}
\end{center}
\caption{\label{Figure3_STM_gap}(Colour online) (a) Mapping ($30\times 30$ nm$^2$) of the SC gap size of a Ca$_{0.83}$La$_{0.17}$Fe$_2$As$_2$ sample. The inset shows a typical STM spectrum. (b) Histogram of the gap size distribution. The inset shows the corresponding average spectrum.} 
\end{figure}

We now turn to the characterization of the SC gap. Despite a high $T_c$ and relatively good photoemission band dispersion features, no SC coherent peak was observed by ARPES in the SC state and our attempts to extract a SC gap using this technique, even using a He discharge lamp with high-energy resolution, were unfortunately unsuccessful. Accordingly, the STM measurements of the SC gap do not exhibit well-defined SC coherent peaks. However,  a small shoulder accompanied with a depression in dI/dV centred at 0 mV bias, as shown in the inset of Fig. \ref{Figure3_STM_gap}(a) for a typical STM spectrum, allows us to estimate the size of a gap-like depression in the density of states as a function of position. The corresponding gap mapping ($30\times 30$ nm$^2$) is displayed in Fig. \ref{Figure3_STM_gap}(a). Large variations of the gap and its random distribution are observed at the nanoscopic scale. This set of data is representative of all sample pieces and all areas we measured, suggesting that the gap is homogeneous at the mesoscopic and macroscopic scales. The histogram of the gap size distribution is shown in Fig. \ref{Figure3_STM_gap}(b). A gaussian fit indicates that the distribution is centred at 7.7 meV, with a half-width at half-maximum of 2.9 meV, which is consistent with the average spectrum given in the inset of Fig. \ref{Figure3_STM_gap}(b). Because of its size and the electron-hole symmetry of the spectra, we argue that this gap is most likely due to superconductivity. Unfortunately, we can neither ascribe this SC gap value to a particular FS pocket nor determine whether it is associated to the SC transition at 20 K or at 42 K. However, it worths mentioning that such gap value would place this system in the strong coupling limit, even if we consider $T_c=42$ K to calculate the $2\Delta/k_BT_c$ ratio. Moreover, we point out that the randomness of the SC gap distribution at the nanoscopic scale and its homogeneity at the mesoscopic scale seem inconsistent with the proposed SC minority phases. Although the origin of the low SC volume fraction remains unclear and further studies are necessary to conclude on this issue, we conjecture that it may be related to the traces of antiferromagnetic band folding observed by ARPES. 

In summary, our combine ARPES and STM data on electron-doped Ca$_{0.83}$La$_{0.17}$Fe$_2$As$_2$ samples indicate the presence of a surface reconstruction due to the dimerization of As surface atoms. This reconstruction put aside, the FS topology of Ca$_{0.83}$La$_{0.17}$Fe$_2$As$_2$ is similar to that of BaFe$_{1.85}$Co$_{0.15}$As$_2$, where $\Gamma$-centred hole pockets and M-centred electron pockets are quasi-nested by the AFM wave vector. Due to low SC volume fraction, no sharp SC coherent peak is detected. Our STM study reveals a gap-like depression in the density of states with large variations that averages to $7.7\pm 2.9$ meV.

\section*{ACKNOWLEDGMENTS}
This ARPES work was supported by grants from CAS (2010Y1JB6), MOST (2010CB923000 and 2011CBA001000), and NSFC (11004232 and 11050110422) of China, as well as by the Sino-Swiss Science and Technology Cooperation (project no. IZLCZ2 138954). The work in Houston is supported in part by US Air Force Office of Scientific Research Contract No. FA9550-09-1-0656, Department of Energy Subcontract No. 4000086706 through ORNL, AFRL Subcontract No. R15901 (CONTACT) through Rice University, the T. L. L. Temple Foundation and the John J. and Rebecca Moores Endowment, and the State of Texas through TCSUH. 

\bibliography{biblio_long}

\begin{thebibliography}{16}%
\makeatletter
\providecommand \@ifxundefined [1]{%
 \@ifx{#1\undefined}
}%
\providecommand \@ifnum [1]{%
 \ifnum #1\expandafter \@firstoftwo
 \else \expandafter \@secondoftwo
 \fi
}%
\providecommand \@ifx [1]{%
 \ifx #1\expandafter \@firstoftwo
 \else \expandafter \@secondoftwo
 \fi
}%
\providecommand \natexlab [1]{#1}%
\providecommand \enquote  [1]{``#1''}%
\providecommand \bibnamefont  [1]{#1}%
\providecommand \bibfnamefont [1]{#1}%
\providecommand \citenamefont [1]{#1}%
\providecommand \href@noop [0]{\@secondoftwo}%
\providecommand \href [0]{\begingroup \@sanitize@url \@href}%
\providecommand \@href[1]{\@@startlink{#1}\@@href}%
\providecommand \@@href[1]{\endgroup#1\@@endlink}%
\providecommand \@sanitize@url [0]{\catcode `\\12\catcode `\$12\catcode
  `\&12\catcode `\#12\catcode `\^12\catcode `\_12\catcode `\%12\relax}%
\providecommand \@@startlink[1]{}%
\providecommand \@@endlink[0]{}%
\providecommand \url  [0]{\begingroup\@sanitize@url \@url }%
\providecommand \@url [1]{\endgroup\@href {#1}{\urlprefix }}%
\providecommand \urlprefix  [0]{URL }%
\providecommand \Eprint [0]{\href }%
\providecommand \doibase [0]{http://dx.doi.org/}%
\providecommand \selectlanguage [0]{\@gobble}%
\providecommand \bibinfo  [0]{\@secondoftwo}%
\providecommand \bibfield  [0]{\@secondoftwo}%
\providecommand \translation [1]{[#1]}%
\providecommand \BibitemOpen [0]{}%
\providecommand \bibitemStop [0]{}%
\providecommand \bibitemNoStop [0]{.\EOS\space}%
\providecommand \EOS [0]{\spacefactor3000\relax}%
\providecommand \BibitemShut  [1]{\csname bibitem#1\endcsname}%
\let\auto@bib@innerbib\@empty
\bibitem [{\citenamefont {{P. Richard, K. Nakayama, T. Sato, M. Neupane, Y.-M.
  Xu, J. H. Bowen, G. F. Chen, J. L. Luo, N. L. Wang, X. Dai, Z. Fang, H. Ding
  and T. Takahashi}}(2010)}]{RichardPRL2010}%
  \BibitemOpen
  \bibfield  {author} {\bibinfo {author} {\bibnamefont {{P. Richard, K.
  Nakayama, T. Sato, M. Neupane, Y.-M. Xu, J. H. Bowen, G. F. Chen, J. L. Luo,
  N. L. Wang, X. Dai, Z. Fang, H. Ding and T. Takahashi}}},\ }\href@noop {}
  {\bibfield  {journal} {\bibinfo  {journal} {Phys. Rev. Lett.}\ }\textbf
  {\bibinfo {volume} {104}},\ \bibinfo {pages} {137001} (\bibinfo {year}
  {2010})}\BibitemShut {NoStop}%
\bibitem [{\citenamefont {{S. A. J. Kimber, A. Kreyssig, Y.-Z. Zhang, H. O.
  Jeschke, R. Valentí, F. Yokaichiya, E. Colombier, J. Yan, T. C. Hansen, T.
  Chatterji, R. J. McQueeney, P. C. Canfield, A. I. Goldman and D. N.
  Argyriou}}(2009)}]{Kimber}%
  \BibitemOpen
  \bibfield  {author} {\bibinfo {author} {\bibnamefont {{S. A. J. Kimber, A.
  Kreyssig, Y.-Z. Zhang, H. O. Jeschke, R. Valentí, F. Yokaichiya, E.
  Colombier, J. Yan, T. C. Hansen, T. Chatterji, R. J. McQueeney, P. C.
  Canfield, A. I. Goldman and D. N. Argyriou}}},\ }\href@noop {} {\bibfield
  {journal} {\bibinfo  {journal} {Nature Materials}\ }\textbf {\bibinfo
  {volume} {8}},\ \bibinfo {pages} {471} (\bibinfo {year} {2009})}\BibitemShut
  {NoStop}%
\bibitem [{\citenamefont {{M. S. Torikachvili, S. L. Bud'ko, N. Ni and P. C.
  Canfield}}(2008)}]{Torikachvili2}%
  \BibitemOpen
  \bibfield  {author} {\bibinfo {author} {\bibnamefont {{M. S. Torikachvili, S.
  L. Bud'ko, N. Ni and P. C. Canfield}}},\ }\href@noop {} {\bibfield  {journal}
  {\bibinfo  {journal} {Phys. Rev. B}\ }\textbf {\bibinfo {volume} {78}},\
  \bibinfo {pages} {104527} (\bibinfo {year} {2008})}\BibitemShut {NoStop}%
\bibitem [{\citenamefont {{G. F. Chen, Z. Li, J. Dong, G. Li, W. Z. Hu, X. D.
  Zhang, X. H. Song, P. Zheng, N. L. Wang and J. L. Luo}}(2008)}]{GF_Chen2}%
  \BibitemOpen
  \bibfield  {author} {\bibinfo {author} {\bibnamefont {{G. F. Chen, Z. Li, J.
  Dong, G. Li, W. Z. Hu, X. D. Zhang, X. H. Song, P. Zheng, N. L. Wang and J.
  L. Luo}}},\ }\href@noop {} {\bibfield  {journal} {\bibinfo  {journal} {Phys.
  Rev. B}\ }\textbf {\bibinfo {volume} {78}},\ \bibinfo {pages} {224512}
  (\bibinfo {year} {2008})}\BibitemShut {NoStop}%
\bibitem [{\citenamefont {{M. Rotter, M. Tegel and D.
  Johrendt}}(2008)}]{Rotter_PRL2008}%
  \BibitemOpen
  \bibfield  {author} {\bibinfo {author} {\bibnamefont {{M. Rotter, M. Tegel
  and D. Johrendt}}},\ }\href@noop {} {\bibfield  {journal} {\bibinfo
  {journal} {Phys. Rev. Lett.}\ }\textbf {\bibinfo {volume} {101}},\ \bibinfo
  {pages} {107006} (\bibinfo {year} {2008})}\BibitemShut {NoStop}%
\bibitem [{\citenamefont {{B. Lv, L. Deng, M. Gooch, F. Wei, Y. Sun, J. K.
  Meen, Y.-Y. Xue, B. Lorenz and C.-W. Chu}}(2011)}]{LvPNAS2011}%
  \BibitemOpen
  \bibfield  {author} {\bibinfo {author} {\bibnamefont {{B. Lv, L. Deng, M.
  Gooch, F. Wei, Y. Sun, J. K. Meen, Y.-Y. Xue, B. Lorenz and C.-W. Chu}}},\
  }\href@noop {} {\bibfield  {journal} {\bibinfo  {journal} {Proc. Natl. Acad.
  Sci. USA}\ }\textbf {\bibinfo {volume} {108}},\ \bibinfo {pages} {15705}
  (\bibinfo {year} {2011})}\BibitemShut {NoStop}%
\bibitem [{\citenamefont {{Z. Gao, Y. Qi, L. Wang, D. Wang, X. Zhang, C. Yao,
  C. Wang and Y. Ma}}(2011)}]{Gao_EPL95}%
  \BibitemOpen
  \bibfield  {author} {\bibinfo {author} {\bibnamefont {{Z. Gao, Y. Qi, L.
  Wang, D. Wang, X. Zhang, C. Yao, C. Wang and Y. Ma}}},\ }\href@noop {}
  {\bibfield  {journal} {\bibinfo  {journal} {Europhys. Lett.}\ }\textbf
  {\bibinfo {volume} {95}},\ \bibinfo {pages} {67002} (\bibinfo {year}
  {2011})}\BibitemShut {NoStop}%
\bibitem [{\citenamefont {{S. R. Saha, N. P. Butch, T. Drye, J. Magill, S.
  Ziemak, K. Kirshenbaum, P. Y. Zavalij, J. W. Lynn and J.
  Paglione}}(2012)}]{Saha_PRB85}%
  \BibitemOpen
  \bibfield  {author} {\bibinfo {author} {\bibnamefont {{S. R. Saha, N. P.
  Butch, T. Drye, J. Magill, S. Ziemak, K. Kirshenbaum, P. Y. Zavalij, J. W.
  Lynn and J. Paglione}}},\ }\href@noop {} {\bibfield  {journal} {\bibinfo
  {journal} {Phys. Rev. B}\ }\textbf {\bibinfo {volume} {85}},\ \bibinfo
  {pages} {024525} (\bibinfo {year} {2012})}\BibitemShut {NoStop}%
\bibitem [{\citenamefont {{P. Richard, T. Sato, K. Nakayama, T. Takahashi and
  H. Ding}}(2011)}]{RichardRoPP2011}%
  \BibitemOpen
  \bibfield  {author} {\bibinfo {author} {\bibnamefont {{P. Richard, T. Sato,
  K. Nakayama, T. Takahashi and H. Ding}}},\ }\href@noop {} {\bibfield
  {journal} {\bibinfo  {journal} {Rep. Prog. Phys.}\ }\textbf {\bibinfo
  {volume} {74}},\ \bibinfo {pages} {124512} (\bibinfo {year}
  {2011})}\BibitemShut {NoStop}%
\bibitem [{\citenamefont {{M. Neupane, P. Richard, Y.-M. Xu, K. Nakayama, T.
  Sato, T. Takahashi, A. V. Federov, G. Xu, X. Dai, Z. Fang, Z. Wang, G.-F.
  Chen, N.-L. Wang, H.-H. Wen and H. Ding}}(2011)}]{Neupane_PRB2011}%
  \BibitemOpen
  \bibfield  {author} {\bibinfo {author} {\bibnamefont {{M. Neupane, P.
  Richard, Y.-M. Xu, K. Nakayama, T. Sato, T. Takahashi, A. V. Federov, G. Xu,
  X. Dai, Z. Fang, Z. Wang, G.-F. Chen, N.-L. Wang, H.-H. Wen and H. Ding}}},\
  }\href@noop {} {\bibfield  {journal} {\bibinfo  {journal} {Phys. Rev. B}\
  }\textbf {\bibinfo {volume} {83}},\ \bibinfo {pages} {094522} (\bibinfo
  {year} {2011})}\BibitemShut {NoStop}%
\bibitem [{\citenamefont {{J. E. Hoffman}}(2011)}]{HoffmanRoPP2011}%
  \BibitemOpen
  \bibfield  {author} {\bibinfo {author} {\bibnamefont {{J. E. Hoffman}}},\
  }\href@noop {} {\bibfield  {journal} {\bibinfo  {journal} {Rep. Prog. Phys.}\
  }\textbf {\bibinfo {volume} {74}},\ \bibinfo {pages} {124513} (\bibinfo
  {year} {2011})}\BibitemShut {NoStop}%
\bibitem [{\citenamefont {{P. Zhang, P. Richard, T. Qian, Y.-M. Xu, X. Dai and
  H. Ding}}(2011)}]{P_Zhang_RSI2011}%
  \BibitemOpen
  \bibfield  {author} {\bibinfo {author} {\bibnamefont {{P. Zhang, P. Richard,
  T. Qian, Y.-M. Xu, X. Dai and H. Ding}}},\ }\href@noop {} {\bibfield
  {journal} {\bibinfo  {journal} {Rev. Sci. Instrum.}\ }\textbf {\bibinfo
  {volume} {82}},\ \bibinfo {pages} {043712} (\bibinfo {year}
  {2011})}\BibitemShut {NoStop}%
\bibitem [{\citenamefont {{A. Damascelli}}(2004)}]{DamascelliPScrypta2004}%
  \BibitemOpen
  \bibfield  {author} {\bibinfo {author} {\bibnamefont {{A. Damascelli}}},\
  }\href@noop {} {\bibfield  {journal} {\bibinfo  {journal} {Phys. Scrypta}\
  }\textbf {\bibinfo {volume} {T109}},\ \bibinfo {pages} {61} (\bibinfo {year}
  {2004})}\BibitemShut {NoStop}%
\bibitem [{\citenamefont {{K. Terashima, Y. Sekiba, J. H. Bowen, K. Nakayama,
  T. Kawahara, T. Sato, P. Richard, Y.-M. Xu, L. J. Li, G. H. Cao, Z.-A. Xu, H.
  Ding and T. Takahashi}}(2009)}]{Terashima_PNAS2009}%
  \BibitemOpen
  \bibfield  {author} {\bibinfo {author} {\bibnamefont {{K. Terashima, Y.
  Sekiba, J. H. Bowen, K. Nakayama, T. Kawahara, T. Sato, P. Richard, Y.-M. Xu,
  L. J. Li, G. H. Cao, Z.-A. Xu, H. Ding and T. Takahashi}}},\ }\href@noop {}
  {\bibfield  {journal} {\bibinfo  {journal} {Proc. Natl. Acad. Sci. USA}\
  }\textbf {\bibinfo {volume} {106}},\ \bibinfo {pages} {7330} (\bibinfo {year}
  {2009})}\BibitemShut {NoStop}%
\bibitem [{\citenamefont {{C. Liu, T. Kondo, N. Ni, A. D. Palczewski, A.
  Bostwick, G. D. Samolyuk, R. Khasanov, M. Shi, E. Rotenberg, S. L. Bud'ko, P.
  C. Canfield and A. Kaminski}}(2009)}]{LiuPRL2009}%
  \BibitemOpen
  \bibfield  {author} {\bibinfo {author} {\bibnamefont {{C. Liu, T. Kondo, N.
  Ni, A. D. Palczewski, A. Bostwick, G. D. Samolyuk, R. Khasanov, M. Shi, E.
  Rotenberg, S. L. Bud'ko, P. C. Canfield and A. Kaminski}}},\ }\href@noop {}
  {\bibfield  {journal} {\bibinfo  {journal} {Phys. Rev. Lett.}\ }\textbf
  {\bibinfo {volume} {102}},\ \bibinfo {pages} {167004} (\bibinfo {year}
  {2009})}\BibitemShut {NoStop}%
\bibitem [{\citenamefont {{Y. Zhang, J.Wei, H.W. Ou, J. F. Zhao, B. Zhou, F.
  Chen, M. Xu, C. He, G.Wu, H. Chen, M. Arita, K. Shimada, H. Namatame, M.
  Taniguchi, X. H. Chen and D. L. Feng}}(2009)}]{Y_Zhang_PRL102}%
  \BibitemOpen
  \bibfield  {author} {\bibinfo {author} {\bibnamefont {{Y. Zhang, J.Wei, H.W.
  Ou, J. F. Zhao, B. Zhou, F. Chen, M. Xu, C. He, G.Wu, H. Chen, M. Arita, K.
  Shimada, H. Namatame, M. Taniguchi, X. H. Chen and D. L. Feng}}},\
  }\href@noop {} {\bibfield  {journal} {\bibinfo  {journal} {Phys. Rev. Lett.}\
  }\textbf {\bibinfo {volume} {102}},\ \bibinfo {pages} {127003} (\bibinfo
  {year} {2009})}\BibitemShut {NoStop}%
\end{thebibliography}%

\end{document}